\documentstyle[twoside,fleqn,espcrc2]{article}

\newcommand{\AmS}{{\protect\the\textfont2
  A\kern-.1667em\lower.5ex\hbox{M}\kern-.125emS}}

\title{
$B\rightarrow K^* \gamma$ decay on APE 
\thanks{Talk presented  by Ph. Boucaud
 \hfill\break 
$^\dagger$ Laboratoire associ\'e au CNRS}
}

\author{      
As.Abada\address{ LPTHE, Universit\'e Paris-XI, 91405 Orsay, France$^\dagger$},
Ph.Boucaud$^{\rm a}$,
M.Crisafulli\address{ Dip. di Fisica, Univ. di Roma \lq La Sapienza\rq},
J.P.Leroy$^{\rm a}$,
V.Lubicz$^{\rm b}$,
G.Martinelli$^{\rm b,}$\address{ Theory Division, CERN, 1211 GENEVA 23, Switzerland.},\\
F. Rapuano$^{\rm b,c}$,
M. Serone$^{\rm b}$,
N.Stella$^{\rm b}$,\\
and \\
A Bartoloni$^{\rm b}$, 
C. Battista$^{\rm b}$, 
S. Cabasino$^{\rm b}$, 
N. Cabibbo\address{ Dip. di Fisica, Univ. di Roma \lq Tor Vergata\rq\  and INFN
, Sezione di Roma II,\\
Via della Ricerca Scientifica 1, I-00133 Roma,Italy .\\}, 
E. Panizzi$^{\rm b}$, 
P.S. Paolucci$^{\rm b}$,
R. Sarno$^{\rm b}$,\\ 
G.M. Todesco$^{\rm b}$, 
M. Torelli$^{\rm b}$, 
P. Vicini$^{\rm b}$.\\
The APE Collaboration 
}

\begin{document}

\begin{abstract}
We present results for the radiative decay $B\rightarrow K^* \gamma$, 
obtained by using the 
Clover action at $\beta = 6.0$ on APE. 
The compatibility between  the scaling laws predicted by the Heavy
Quark Effective Theory (HQET) and  pole dominance is discussed.
The final result depends crucially on the assumed $q^2$-dependence 
of the form factors.
\end{abstract}

\maketitle

\section{Introduction}
The hadronic matrix element  which governs the 
radiative decay $B\rightarrow K^* \gamma$ is 
parametrised in terms
of three form factors:
\begin{eqnarray}
& &\langle  K^*_r(\eta,k) \vert  J_\mu  
\vert B(p)\rangle= \nonumber \\
& & \quad\quad C_1^\mu T_1(q^2)+iC_2^\mu T_2(q^2)+iC_3^\mu T_3(q^2),
\label{parametrization}
\end{eqnarray}
where
\begin{eqnarray}
\label{ci}
C_1^\mu\ =\ 2\epsilon^{\mu\alpha\rho\sigma}\eta_r(k)_\alpha p_\rho
k_\sigma, \nonumber \\
C_2^\mu\ =\ \eta_r(k)(M_B^2-M_{K^*}^2)-(\eta_r(k).q)(p+k)^\mu,
\nonumber  \\ 
C_3^\mu\ =\ \eta_r(k).q\bigl( q^\mu-{q^2\over
M_B^2-M_{K^*}^2}(p+k)^\mu \bigr).  
\end{eqnarray}
and $J_\mu = \bar s \sigma_{\mu\nu}\frac{1\ +\ \gamma_5} {2} q^\nu b$;
$\eta$ is the  polarization vector of the $K^*$ and $q$ the momentum transfer.
When the emitted photon is real, 
$T_3$ does not contribute to the physical rate and 
${T_1(0)=T_2(0)}$. 
At $q^2=0$, the physics of this decay is thus described by only one form factor,
$T_1$.
The feasibility of the lattice approach has been demonstrated first
by the work of Bernard et al. \cite{hsieh}.

\section{Scaling laws and $q^2$ dependence of the form factors}
In order to obtain the form factors at the physical point,
we need to extrapolate  both to large meson  masses and small values of  $q^2$. 
The final results critically depend
on the assumptions made on the $q^2$- and heavy mass-dependence. 
At fixed $\vec p_{K^*}$, with  $\vert \vec p_{K^*}\vert\ll  M_B$ in the
B-meson  rest frame, the following scaling laws can be derived \cite{wise}: 
\begin{eqnarray}
\label{scala} 
\frac{T_1}{\sqrt{M_B}}& = & \gamma _{1}\times \left( 1+\frac{\delta _{1}}{M_B}
+ \dots \right) \nonumber \\  
T_2\sqrt{M_B}& = & \gamma _2\times \left( 1+\frac{\delta _2}{M_B}
+ \dots \right)
\end{eqnarray}
which are valid up to logarithmic corrections.
On the other hand,   ``scaling" laws
for the form factors at $q^2=0$ can only be found by using extra assumptions
for their $q^2$ dependence.  This procedure is  acceptable,
provided the ``scaling" laws derived in this way  respect the exact condition
$T_1(0)=T_2(0)$. 
This is a non-trivial constraint: 
the $q^2$ behaviour of $T_1$ and $T_2$
has to compensate for the different mass dependence of the two form factors 
near the zero recoil point given in eq.(\ref{scala}). For example, the
popular assumption of pole dominance for both $T_1$ and $T_2$  would give
that $T_1(0) \sim M_B^{-1/2}$ and $T_2(0) \sim M_B^{-3/2}$, 
which is inconsistent.
\par
The assumptions on  the $q^2$-dependence of the form factors can be
tested directly on the numerical results,
although only in a small domain of  momenta.

\section{Lattice set-up}
The numerical simulation was performed on the 
6.4 Gigaflops version of the APE machine, 
at $\beta \ = \ 6.0$, on a $18^3\times 64$ lattice, 
using the SW-Clover action \cite{clover}\ in the quenched approximation.
The results have been obtained from a sample of 170 gauge configurations and
the statistical errors  estimated by a jackknife  procedure 
with a decimation of 10 configurations from the total set.
For each configuration we have computed the quark propagators for seven
values of the Wilson hopping parameter $K_W$, corresponding to
\lq\lq heavy" quarks, $K_H=0.1150$, $0.1200$, $0.1250$, $0.1330$,
and \lq\lq light" quarks, $K_L=0.1425$, $0.1432$ and $0.1440$. 
Due to memory limitations, the propagators are ``thinned".  
The matrix elements have been computed for an initial meson
at rest and a final vector meson  with momentum $\vec p_{K^*}$.
We have taken $\vec p_{K^*} =  2 \pi / (La)$ 
 $(0,0,0)$, $(1,0,0)$, $(1,1,0)$, $(1,1,1)$, and  $(2,0,0)$, where $L$  
is the spatial extension of the lattice.
The initial (final) meson was created (annihilated)
by  using a pseudoscalar (local vector)  density 
inserted at a time $t_B/a=28$ ($t_{K^*}=0$), and  
we have varied the time position of the current 
in the interval $t_J/a=10-14$. 
Two procedures, denoted by ``ratio" and ``analytic" in the tables, 
have been used to extract the plateaux: the three point functions
are divided either by the numerical  two point functions (``ratio")
or  by an analytical expression (``analytic"). 
Details can be found in ref \cite{abada2}.

\section{Results}
\begin{figure}[htb]
\vspace{9pt}
\vskip 2.8truecm
\includegraphics{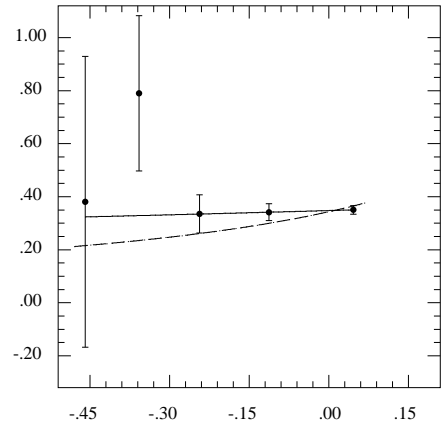}
\vskip 1.3truecm
\caption{$T_2(q^2)$ as a function of $q^2$ for $K_H =.1330$. 
The curves show the pole dominance with either the lattice 
axial pole mass (dashed line)
or with a free pole mass (full line).}
\label{fig:T2}
\end{figure}

From our data, if we assume a pole dominance behaviour for $T_2$, 
the mass extracted from the fits with the pole mass as a free parameter 
is larger than the  mass obtained from the axial two-point  correlation 
functions.
As a consequence  $T_2(q^2)$ is flatter than predicted by pole dominance
(see fig. \ref{fig:T2}). 
The values found 
in  this way  at $q^2=0$  are reported in 
table \ref{tab:TunTdeux} as $T_2^{free}(0)$.
\vskip -.6truecm
In the case of $T_1$, the absence of data at $q^2_{max}$ and the large 
errors in the data at high momenta make it difficult to test directly 
the validity of the pole dominance hypothesis. 
We can use the value for $T_1(q^2=0)$ obtained from the 
the condition $T_1(0)=T_2^{free}(0)$, together with the point at 
$\vec p_{K^*} =  2 \pi / (La)$ $(1,0,0)$
in a fit of $T_1$ 
to  a pole dominance behaviour; the
pole mass determined along this way  for $T_1$
is compatible, though with large errors, with the mass of the corresponding 
lattice vector meson. More data (i.e. with a moving  $B$ meson)
are needed to test  this point more accurately.
As a matter of comparison, we give  in 
 table \ref{tab:TunTdeux}
 the  values for  $T_1^{pole}(q^2=0)$ ($T_2^{pole}(q^2=0)$)
obtained under the assumption of 
 a pole dominance with the lattice vector (axial) meson mass.
Although the quality of our data is not accurate enough  
to draw a definite conclusion, 
they suggest that  assuming  $T_2$ flatter than  pole dominance 
and $T_1$ following pole dominance   gives a good description of our data. 
We call this option $m^{-1/2}$-scaling.
In fig.  \ref{fig:T2}, the curve corresponding to 
the pole dominance for $T_2$  is also given  ($m^{-3/2}$-scaling);
this scaling law would follow from a dipolar $q^2$-dependence for $T_1$.
We take the two possibilities,  $m^{-1/2}$- and $m^{-3/2}$-scaling,
as representatives of a whole  class of possible ``scaling" laws.
\vskip -.6truecm
\begin{table}[hbt]
\caption{Form factors at  $q^2=0$ extrapolated
to the strange quark, assuming independence on the spectator quark. 
``ratio" and ``analytic" are explained in the text.
$T_1^{pole}$ and $T_2^{pole}$ are computed with the appropriate 
lattice meson mass for the pole dominance, $T_2^{free}$ with the pole mass
as a free parameter.}
\label{tab:TunTdeux}
\centering
\begin{tabular} {|r|c|c|}
\hline
              & {\rm ratio}   & {\rm analytic }   \\
\hline 
$T_1^{pole}(0) \kappa_h$ = .1150   & .286(35) & .297(34) \\
$T_2^{free}(0) \kappa_h$ = .1150   & .280(52) & .301(56) \\
$T_2^{pole}(0) \kappa_h$ = .1150   & .238(17) & .242(17) \\
                \hline
$T_1^{pole}(0) \kappa_h$ = .1200   & .298(33) & .309(37) \\
$T_2^{free}(0) \kappa_h$ = .1200   & .293(40) & .309(40) \\
$T_2^{pole}(0) \kappa_h$ = .1200   & .262(16) & .265(17) \\
              \hline
$T_1^{pole}(0) \kappa_h$ = .1250   & .311(32) & .322(31) \\
$T_2^{free}(0) \kappa_h$ = .1250   & .310(30) & .320(28) \\
$T_2^{pole}(0) \kappa_h$ = .1250   & .288(17) & .292(17) \\
                \hline               
$T_1^{pole}(0) \kappa_h$ = .1330   & .331(31) & .339(30) \\
$T_2^{free}(0) \kappa_h$ = .1330   & .345(19) & .348(18) \\
$T_2^{pole}(0) \kappa_h$ = .1330   & .340(19) & .343(19) \\
\hline
\end{tabular}
\end{table}

The extrapolation to the physical region (i.e.  the $B$ mass)  
is performed following
these  two hypothesis, $m^{-1/2}$- and $m^{-3/2}$-scaling 
with linear and quadratic fits. 
The results are presented in   table \ref{tab:TatB}. 
We give also $\delta_1$, the coefficient of the $1/M$ corrections in the
linear fit. 
It should be noted that in the $m^{-1/2}$-scaling case,  the $1/M$ corrections
are smaller and the extrapolated value is less affected by  
the adjunction of a quadratic term than in the  $m^{-3/2}$-scaling hypothesis. 
\begin{table}[hbt]
\caption{The form factors at $q^2=0$ extrapolated to  the physical $B$ mass.
 $\delta_1$ is the coefficient of the $1/M$ correction.}
\label{tab:TatB}
\centering
\begin{tabular} {|r|c|c|}
\hline
              & {\rm ratio}   & {\rm analytic }   \\
\hline 
$T_1^{pole}(0)$ fit $m^{-{1\over2}}$ lin. &  .203(28)  &  .213(27)   \\
$T_1^{pole}(0)$ fit $m^{-{1\over2}}$ quad.&  .191(40)  &   .200(40)      \\
$\delta_1$ (MeV)  &  310(109)  &   339(97)      \\
              \hline
$T_1^{pole}(0)$ fit $m^{-{3\over2}}$ lin. &  .102(11)  &  .106(12)      \\
$T_1^{pole}(0)$ fit $m^{-{3\over2}}$ quad.&  .135(20)  &  .140(21)      \\
$\delta_1 $(MeV) &  871(34)  &   879(34)      \\
              \hline
$T_2^{pole}(0)$ fit $m^{-{3\over2}}$ lin. &  .082(7)  &  .083(7)      \\
$T_2^{pole}(0)$ fit $m^{-{3\over2}}$ quad.&  .091(12) &  .092(13)     \\
$\delta_1 $(MeV) & 735(51)   &    737(54)     \\
\hline
\end{tabular}
\end{table}

As a consistency check, we can first extrapolate $T_1$ to the $B$,  
at (small) fixed momentum, following the scaling law of eq. \ref{scala}.
We have used $\vec p_{K^*} =  2 \pi / (La)$ $(1,0,0)$. Then from a pole 
dominance
with the physical vector meson mass ($M_V \sim 5.4$ GeV) we get for  the value
of $T_1(0)$ the results   0.192(44) and .200(44) 
for the ``ratio" and ``analytic" 
methods respectively, in agreement with the results in table \ref{tab:TatB}
($m^{-1/2}$-scaling). The same game can be played with $T_2(q^2_{max})$; 
the results in this case are $T_2(q^2_{max},B)= .217(15)$ and 
$T_2(q^2=0,B)= .090(6)$ for 
a  value of $M_A \sim 5.7$ GeV for the axial pole mass; this is
in agreement with the results in table \ref{tab:TatB} 
for the $m^{-3/2}$-scaling.

From table \ref{tab:TatB}, we quote:
\begin{eqnarray*}
T_1^{pole}(0) = .196(45) \ \ \ (m^{-1/2}\ \ {\rm scaling })\\
T_2^{pole}(0) = .090(15)         \ \ \ (m^{-3/2} \ \ {\rm  scaling })\\
\end{eqnarray*}

Clearly, the final result depends crucially on the 
assumption made for the $q^2$-dependence. 
Given the statistical errors, the
systematic uncertainty in the extraction of the form factors, the
effects of $O(a)$ terms  and the limited range in $q^2$ and masses, 
the study of the $q^2$- and mass-dependence of the form factors, 
remains a crucial challenge for lattice calculations.

\end{document}